# Broadband model-based inversion enables optoacoustic microscopy beyond the acoustic diffraction limit

Weiye Li[†], Urs A. T. Hofmann[†], Johannes Rebling, Ali Ozbek, Yuxiang Gong, Daniel Razansky, and Xose Luis Dean-Ben[*]

*Institute for Biomedical Engineering and Institute of Pharmacology and Toxicology, University of Zurich and ETH Zurich, Switzerland*

*Corresponding author: xl.deanben@uzh.ch*

[†]*These authors contributed equally to this work*

## Abstract

Acoustic-resolution optoacoustic microscopy (AR-OAM) retrieves anatomical and functional contrast from living tissues at depths not reachable with optical microscopy. The imaging performance of AR-OAM has been advanced with image reconstruction algorithms providing high lateral resolution ultimately limited by acoustic diffraction. In this work, we suggest a new model-based framework efficiently exploiting scanning symmetries for high-resolution reconstruction of AR-OAM images. The model accurately accounts for the spatial impulse response and large detection bandwidth of a spherical polyvinylidene difluoride sensor, which facilitates significantly outperforming synthetic aperture focusing technique commonly employed in AR-OAM image reconstruction in terms of image contrast and resolution. Furthermore, reconstructions based on L1-norm regularization enabled resolving structures indistinguishable with other methods, which was confirmed by numerical simulations as well as phantom and *in vivo* experiments. The achieved performance demonstrates the applicability of AR-OAM as a super-resolution imaging method capable of breaking through the limits imposed by acoustic diffraction, thus opening unprecedented capabilities for the microscopic interrogation of optically opaque tissues in preclinical and clinical studies.

## Introduction

Advanced acoustic inversion methods are essential to optimize the achievable resolution, contrast, and overall performance of optoacoustic (OA, photoacoustic) imaging systems operating at depths not reachable with optical microscopy (>1 mm in biological tissues). The type and location of ultrasound (US) sensor(s) mainly determine the imaging performance in this depth range, thus are generally tailored for the biomedical application of interest [1, 2]. For example, OA tomography based on concave piezoelectric arrays has been shown to capitalize on a large angular coverage to render accurate images of arbitrarily-oriented vascular networks [3-5]. However, although tomographic microscopic imaging has been enabled with high-resolution arrays [6], deep-tissue OA microscopy (OAM) is typically performed by raster-scanning a single-element transducer [7-12]. This approach, generally known as acoustic-resolution OAM (AR-OAM), represents one of the most common OA embodiments and is increasingly being used in preclinical and clinical studies [13, 14]. Important advantages of AR-OAM over tomographic OA methods are a significantly reduced system complexity, lower cost laser excitation, and simpler US sensor technology. More importantly, AR-OAM is particularly suitable for exploiting the multiscale OA imaging capabilities. For example, centimeter-scale depths have been achieved by using transducers with detection bandwidths of a few MHz [7, 15], while ultra-wideband detectors have been used e.g. in raster scan OA mesoscopy (RSOM) to image at depths around 2 mm-4 mm [13, 16]. Multiple scales can additionally be covered simultaneously e.g. via coaxial light focusing in hybrid-focus OAM [17, 18] or with a dual-element transducer in quad-mode OAM [19].

The importance of AR-OAM further fosters ongoing efforts on the optimization of reconstruction methods. Diffraction-limited lateral resolution can be achieved around the focal plane by simply superimposing time-resolved (A-mode) OA signals. However, beamforming algorithms are generally used to avoid out-of-focus resolution degradation. The synthetic aperture focusing technique (SAFT), also used in radar, sonar, and biomedical US, has become the standard image reconstruction method in AR-OAM [10]. SAFT was introduced in AR-OAM as the virtual detector (VD) concept [20], and is generally implemented using the delay-and-sum (DAS) technique in the time domain [21]. More elaborated approaches based on weighting factors in the time or frequency domains [22, 23] or on delay-multiply-and-sum (DMAS) methods [24, 25] have further been shown to provide an enhanced performance. SAFT relies on the heuristic principle of signal superposition via constructive interference. However, OA DAS may alternatively be regarded as a discrete implementation of the filtered back-projection algorithm, derived from the time-domain OA forward model for short-pulsed excitation [26]. Such a model represents a solid mathematical foundation enabling the development of more accurate acoustic inversion methods. Thus, algorithms based on time reversal techniques [27, 28], model-based (iterative) inversion algorithms [29-31] or hybrid-domain modeling of a focused transducer [28] have been suggested for AR-OAM. Model-based reconstruction in the time domain is arguably the most accurate approach to account for the finite size and shape of US sensors [32-34]. Additionally, the performance of iterative inversion can be enhanced by properly choosing constraints or regularization terms [35-40]. However, implementation of this method for the dense grid of voxels required for high-resolution volumetric imaging remains challenging due to high computational cost, even when considering efficient parallel implementations in graphics processing units (GPUs) [41, 42].

Herein, we introduce a model-based reconstruction framework that efficiently exploits the scanning symmetries in AR-OAM to enable GPU-based, high-resolution, volumetric reconstructions. The underlying time-domain OA forward model accurately accounts for the large bandwidth of the OA signals collected with a spherical polyvinylidene difluoride (PVDF) transducer. Discretization of the transducer surface additionally enables modeling its spatial impulse response. Model-based inversion is shown to significantly enhance the resolution achieved with SAFT even beyond the acoustic diffraction limit. The suggested approach is generally applicable in any AR-OAM system.

Results

AR-OAM acquires volumetric images in a raster-scan acquisition protocol schematically depicted in Fig. 1a. The Nyquist criterion should be fulfilled by selecting a scanning step smaller than half of the expected lateral resolution. The lateral resolution in AR-OAM is considered to be limited by acoustic diffraction to $0.71 \lambda/NA$ [43], where $\lambda$ and $NA$ are the central wavelength and numerical aperture of the US transducer, respectively. Typical lateral resolutions are in the range of 50 μm, thus scanning steps in the order of a few tens of micrometers are generally needed. The suggested model-based framework defines time-resolved OA signals corresponding to each pair of voxel and scanning point, where the lateral positions of the voxels of the reconstruction grid match the x-y-scanning positions (see methods for a detailed description). For a raster-scan configuration, the model is translational-symmetric — signals are identical if the relative position between the voxel and scanning point is maintained (Fig. 1b). A spherical US transducer further defines axial symmetries with respect to the acoustic axis (Fig. 1b). The OA forward model enables accurate modeling of the impulse response (IR) of the transducer. The IR is commonly split into two terms, namely the electrical impulse response (EIR) defining the effective detection bandwidth, and the spatial impulse response (SIR) accounting for acoustic diffraction effects associated to the finite size of the sensing area [44]. A transducer consisting of a PVDF film featuring broad detection bandwidth (~100 MHz) was used herein [11] (Fig. 1c, see methods for a detailed description), and high axial resolution (~14 μm) has been achieved [11]. However, the spatially-dependent SIR of the spherically-focused, axially-symmetric transducer strongly limits the lateral resolution, especially if standard image reconstruction methods (SAFT and its variants) are employed. In the suggested model-based framework, the SIR for each voxel was estimated by superimposing the corresponding time-resolved OA signals for a set of surface elements (~1000) covering the active sensing area (Fig. 1c). This is equivalent to adding up the OA forward model corresponding to each surface element (see methods for a detailed description). Considering the scanning symmetries, the information in the model-matrix is fully contained in the sub-matrix corresponding to an individual scanning point. Such sub-matrix is sparse and was further set to zero for voxels where the transducer sensitivity drops significantly, thus could be stored in the memory of state-of-the-art GPUs (1.3 GB for half a million voxels and 837 time points, see methods for details). The memory required to store the OA forward model is invariant to the arbitrarily-large raw signal volume, which allows storing the OA signals, the model and the reconstructed image simultaneously on a GPU. Thereby, large volumes can be reconstructed at high-resolution in a reasonable time (~12 minutes). The amplitude spectral density of the time-resolved signals of the OA forward model (columns of the model matrix) reveals the expected diffraction patterns for different frequencies, while their amplitude in the time domain defines a transducer sensitivity field matching the expected diffraction-limited resolution for broadband OA signals (Fig. 1d). The capability to encode frequency-dependent information empowers the suggested model-based framework with unique capabilities to reconstruct absorbers with different sizes emitting OA waves in different frequency bands, clearly outperforming alternative approaches e.g. based on superimposing signals or SAFT (Fig. 1e).

Mathematically, model-based reconstruction minimizes an energy functional consisting of a data fidelity term driving the solution towards the observed data and a regularization term that stabilizes the inversion and includes prior information [45]. Reconstruction of AR-OAM images is challenged by the limited-view scanning geometry and the associated ill-posed nature of the inverse problem [46]. Regularization hence turns essential for computing a credible approximation of the light absorption distribution. Tikhonov regularization, based on the L2 norm, is the classical method for noise-robust image reconstruction and represents the most probable solution given the raw signal and a prior Gaussian distribution of measurement noise. The model-based algorithm based on an L2 regularization term (MB-L2, see methods for details) enabled enhancing the lateral resolution and contrast of AR-OAM images at out-of-focus locations compared to SAFT (Fig. 1e). However, reconstruction of closely separated, simulated absorbers at the focal plane indicate that the lateral resolution is still limited by acoustic diffraction (Figs. 2a and 2b). This is expected considering that MB-L2 is a linear inversion procedure equivalent to time-reversed propagation of US waves [47]. Alternatively, regularization terms based on the L1 norm can be used to promote sparse solutions in a given domain. Particularly, a L1-based regularization term in the image domain translates to most voxels having zero (or close to zero) value. Simulation results with relatively separated absorbers indicated that the full width at half maximum (FWHM) of the absorbers reconstructed with the model-based algorithm with L1 regularization (MB-L1, see methods for details) is considerably smaller (13 μm) than the FWHM achieved with SAFT and MB-L2 (41 μm, Figs. 2a and 2b). This could be erroneously interpreted as a narrower point spread function (PSF) of the imaging system enabling higher resolution. However, MB-L1 is a non-linear inversion procedure and hence a PSF cannot be defined. Instead, the achievable resolution must be estimated from the reconstructed image of closely separated sources. In this regard, undistinguishable absorbers in the images obtained with SAFT and with MB-L2 could be separated with MB-L1, thus effectively breaking through the acoustic diffraction barrier (Figs. 2a and 2b). Experimental results with an agar phantom embedding a ~20 μm microsphere slightly out-of-focus corroborated the simulation results (Figs. 2c and 2d, see methods for a detailed description). The FWHM of the reconstructed sphere was significantly reduced with MB-L1 compared to SAFT and MB-L2, while it was also possible to distinguish two closely separated spheres simulated by shifting the scanning position and adding up the corresponding OA signals (Figs. 2c and 2d, see methods for a detailed description). Additional experiments with an agar phantom embedding carbon fibers with 7 μm diameter further validated the enhanced resolution achieved with MB-L1 (Figs. 2e and 2f). The width of the reconstructed fibers was significantly reduced with MB-L1. More importantly, MB-L1 enabled resolving crossing fibers at points where SAFT and MB-L2 failed (Fig. 2f).

The accurate OA forward model turns essential for precise reconstruction of more complex structures in living organisms. Proper selection of the regularization term is also highly important. Most commonly, OA visualizes multi-scale vascular networks in mammalian tissues with no dominant components in space or frequency domain, which challenges defining information on the images a priori. A raster-scan with 25 μm step across the back of a CD-1 mouse (Fig. 3a, see methods for a detailed description) enabled comparing the *in vivo* imaging performance of different reconstruction approaches. SAFT, MB-L2, and MB-L1 successfully visualized microvascular structures in this region, although clear differences were observed in image quality and content (Fig. 3b). A smoothing effect was observed in the blood vessels reconstructed with SAFT, which appears to degrade the achievable resolution. The vascular network at different depths is clearly better resolved in the MB-L2 image, including branches not visible with SAFT. The sharpening effect of MB-L1 compared to MB-L2 was similar to the observations in simulations and phantom experiments. The width of the reconstructed vessels is clearly narrowed compared to the other two reconstruction techniques. However, this appears to reduce the available information in the image. Many of the branches visible in the MB-L2 image are not seen with MB-L1, arguably due to the sparsity constraint. The achievable resolution with different approaches can be more clearly compared in a three-dimensional view of the central part of the scan (Fig. 3c). Individual vessels and branches in MB-L1 and MB-L2 images are shown to be more clearly defined than in superimposed MB-L2 and SAFT images, respectively. A comparison of the image profiles for a region close to a branching point corroborates that

MB-L1 can break through the resolution limit of the other two methods (Fig. 3d). The respective image content was better assessed by quantifying the amount of vessels in the binarized images with an automatic vessel segmentation and analysis (AVSA) algorithm [48] (Fig. 3e, see methods for a detailed description). Statistical analysis of the number of detected vessels from four different mouse back datasets showed that, on average, the number of vessels in the MB-L2 image is higher by almost two-fold with respect to SAFT and MB-L1 images (Fig. 3f). Overall, the suggested model-based reconstruction framework significantly enhanced the performance of SAFT *in vivo*, while the regularization choice establishes a trade-off between achievable resolution and visibility of vascular networks.

## Discussion

Numerical simulations as well as phantom and *in vivo* experiments demonstrated the enhanced performance of the suggested model-based framework with respect to SAFT as a standard AR-OAM reconstruction technique. This is in agreement with what has been shown in several OA tomographic configurations [38, 49-51]. However, AR-OAM aims at a significantly higher resolution than OA tomography combined with large FOVs. The implementation of volumetric model-based iterative inversion methods is thereby significantly challenged by the fact that the full model matrix, corresponding to the linear operator mapping the initial pressure distribution (OA image) into measured pressure signals, becomes too large to be stored in GPU or even CPU memory. Alternatively, the operations involving the model matrix can be calculated on-the-fly during the inversion procedure, but this has a high computational cost particularly when accounting for the SIR [32, 41]. Herein, we exploited the translational and axial symmetries of the model for the spherically focused transducer within the scanning plane and its limited lateral sensitivity, i.e., absorbers with large lateral distance from the transducer position barely affect measured pressures. The model matrix was "compressed" into a sub-matrix corresponding to a single scanning position and the voxel grids sufficiently close to it. The size of this sub-matrix is defined by the lateral sensitivity, the desired imaging depth, and the number of time instances, i.e., it is invariant with respect to an arbitrarily-large reconstruction grid. Therefore, the suggested model-based reconstruction framework is expected to be generally applicable in any AR-OAM system.

The advantages of model-based inversion stem from the fact that it considers the broadband nature of OA signals. This is fundamentally different to other approaches based on weighting the signals with an approximated transducer sensitivity field, which generally changes for different OA sources due to the frequency dependence of the diffraction-limited acoustic focusing. We have shown in numerical simulations that it is possible to quantitatively reconstruct absorbers of different sizes located at different depths, and further improve the lateral resolution at points outside the depth-of-focus. Modifications of the SAFT method employed for comparison can potentially provide more quantitative results than the standard implementation [52, 53]. However, these modifications are generally heuristically defined, while the suggested framework has a solid mathematical basis on the OA forward model, hence is generally expected to provide more accurate and robust results. In the current implementation, the model matrix was estimated theoretically by assuming that the collected signal is proportional to the integral of the acoustic pressure distribution at the surface of the transducer, which represents a valid approximation for the curved thin PVDF film employed. However, the computation of the model matrix e.g. for transducers based on acoustic lenses or air-coupled transducers is more complex as it involves modeling acoustic interfaces [54, 55]. Also, the finite detection bandwidth influenced by both the PVDF film thickness and the amplification electronics must be considered for a better accuracy. Thereby, a theoretical calculation of the model-matrix may not always be possible. However, the suggested reconstruction methodology can alternatively be used by experimentally estimating the model-matrix for a transducer position e.g. by raster-scanning a small particle or a light beam in the volume of interest [56]. Other parameters can also be considered in the model, such as the light fluence distribution or acoustic attenuation effects, which may lead to further improvements in quantification and spatial resolution [57, 58].

Acoustic diffraction has been considered as a limit for the achievable resolution in OA, similar to light diffraction in optical microscopy [59]. Super-resolution methods have massively impacted life sciences by smartly overcoming the optical diffraction limit [60]. The development of similar approaches enabling breaking through the acoustic diffraction barrier at depths not reachable with optical microscopy is also expected to highly impact the biomedical imaging field and boost the applicability of OA as a research and clinical tool. Well-established optical super-resolution methods can be taken as a reference for this purpose, although fundamental differences exist in physical principles and instrumentation. Recently, localization OA tomography (LOT) exploited the basic principle of photoactivated localization microscopy (PALM) to image beyond the resolution limit imposed by acoustic diffraction [9, 61-64]. The super-resolution method suggested herein (MB-L1) has some analogies with stimulated emission depletion (STED) microscopy as it is based on a raster-scanning protocol where a non-linear response is induced. The non-linearity is however associated with the reconstruction procedure, hence MB-L1 also has similarities with super-resolution methods based on multiple observations with sub-pixel displacements and/or sparse image representations [65, 66]. The use of a L1 regularization term has previously been shown to reduce the width of reconstructed structures in OA tomography [67, 68]. Herein, we have shown that the suggested OA forward model is sufficiently accurate to better resolve closely separated sources – branching vascular structures – *in vivo*, which effectively demonstrates its performance as a super-resolution biomedical imaging tool. The methodology introduced herein further establishes a basic framework for further improvements e.g. by tuning regularization terms, scanning step, or voxel size.

In conclusion, we expect the introduced model-based reconstruction framework to significantly enhance the performance of AR-OAM and enable the visualization of previously unresolvable structures beyond the acoustic diffraction barrier. The fact that the forward model accurately accounts for the transducer spatial response and the broadband nature of pressure waves indicates that model-based inversion arguably represents the most accurate OA image reconstruction method. Considering that image reconstruction can be performed with low memory requirements and in a relatively short time, we anticipate that it may become the method of choice in widely-used AR-OAM systems.

## Methods

### Forward modeling

Excitation of OA signals in AR-OAM is achieved with nanosecond laser pulses fulfilling thermal and stress confinement conditions. In an acoustically-uniform and non-attenuating medium, the time-resolved pressure signal $p(\boldsymbol{x}, t)$ at a point $\boldsymbol{x}$ is given as a function of absorbed optical energy density distribution $H(\boldsymbol{x}')$ as [69]

$$p(\mathbf{x}, t) = \frac{\Gamma}{4\pi c} \frac{\partial}{\partial t} \int_{S'(t)} \frac{H(\mathbf{x}')}{|\mathbf{x} - \mathbf{x}'|} dS'(t), \tag{1}$$

where $\Gamma$ is the Grueneisen parameter, $c$ is the speed of sound, $\mathbf{x}'$ is a point where light is absorbed, and $S'(t)$ is a spherical surface defined as $|\mathbf{x} - \mathbf{x}'| = ct$. The forward model in Eq. 1 was discretized with a two-step procedure — the time derivative was discretized using the finite difference method, and the surface integral was discretized using trilinear interpolation of neighboring voxel values [33]. Assuming a constant $\Gamma$ and $c$, the discrete forward model is then expressed (in arbitrary units) as the linear system of equations

$$\mathbf{p} = \mathbf{AH}, \tag{2}$$

where $\mathbf{p}$ is a vector of pressure signals at all scanning positions, $\mathbf{A}$ is the model matrix and $\mathbf{H}$ is a vector of voxel values representing the absorbed energy density in the region of interest (ROI). The model matrix $\mathbf{A}$ describes the OA response to unit absorption at the voxel grid defined to cover the absorption distribution domain, and can be used to model the signals of a point detector with infinite bandwidth. More specifically, each column vector in $\mathbf{A}$ corresponds to one pair of voxel and scanning point. The OA signal collected by a finite-size transducer can be approximated by integrating pressure waves over the transducer surface [70]. This approximation represents the spatial impulse response (SIR) of the transducer, which can be modeled by dividing the transducer surface into sub-elements with position $x_j$ and area $\Delta x_j$. The vector $\mathbf{s}$ of signals collected by the transducer at all scanning positions is then expressed as

$$\mathbf{s} = \sum_{j=1}^{M} \mathbf{p}_{x_j} \Delta x_j, \tag{3}$$

where $M$ is the number of sub-elements covering the transducer surface (Fig. 1c). Combining Eqs. 2 and 3, the discrete forward model becomes

$$\mathbf{s} = \sum_{j=1}^{M} \mathbf{A}_j \Delta x_j \mathbf{H} = \mathbf{A}_s \mathbf{H}. \tag{4}$$

Thereby, the model matrix $\mathbf{A}_s$ accounting for the SIR of the transducer is estimated as a weighted sum of $M$ model matrices corresponding to $M$ sub-elements.

Efficient reconstruction based on scanning symmetries

Model-based (MB) reconstruction involves minimizing an energy functional

$$\mathbf{H} = argmin_{\mathbf{H}} ||\mathbf{s}_m - \mathbf{A}_s \mathbf{H}||_2^2 + \lambda R(\mathbf{H}), \tag{5}$$

where $\mathbf{s}_m$ is the vector of measured signals, $R(\mathbf{H})$ is regularization term and $\lambda$ is the regularization parameter controlling the trade-off between regularization and data fidelity terms. The importance of $R(\mathbf{H})$ is two-fold — it makes the optimization procedure more robust to noise, and it incorporates a-priori knowledge of the reconstructed image. Herein, we considered the standard Tikhonov regularization based on the L2 norm, where $R(\mathbf{H})$ is given by

$$R(\boldsymbol{H}) = ||\boldsymbol{H}||_2^2. \tag{6}$$

We refer to MB reconstruction with Tikhonov regularization as the MB-L2 method. We also considered L1 regularization based on the L1 norm, where $R(\boldsymbol{H})$ is given by

$$R(\boldsymbol{H}) = ||\boldsymbol{H}||_1. \tag{7}$$

We refer to MB reconstruction with L1 regularization as the MB-L1 method. The solution of Eq. 5 was calculated with the LSQR method [71] for MB-L2 and with the FISTA method [72] for MB-L1.

In both LSQR and FISTA algorithms, the most computationally demanding operations are the matrix-vector multiplications $\boldsymbol{A}_s \boldsymbol{u}$ and $\boldsymbol{A}_s^T \boldsymbol{v}$, where $\boldsymbol{u}$ and $\boldsymbol{v}$ are updated in each iteration. The computational burden originates from the large-scale nature of AR-OAM. This leads to a huge model matrix covering each voxel and scanning position in the entire ROI, even if it is stored in sparse format [41]. However, since the spherically focused transducer barely senses voxels sufficiently distant in the lateral direction, the corresponding model matrix values were approximated as 0, i.e., the model matrix only covers a 1 mm lateral extent (Fig. 1d) to reconstruct arbitrarily large ROI. The depth extent of model matrix scales with the collected dataset. Also, given that the transducer sensitivity field is translational-symmetric, the model vector is equivalent if the relative position between voxel and scanning point is maintained. Therefore, it was sufficient to generate the model matrix for a single scanning position, as it includes all unique lateral distances between pairs of voxel and scanning point. The "compressed" model matrix can then be pre-calculated in a reasonable time and stored in a memory-efficient way. To perform the matrix-vector multiplications, translational symmetries were considered to find the correspondences between voxels and model vectors (Fig. 1b). Specifically, we maintained two coordinate systems — an absolute coordinate system to index over the voxel grid, and a relative coordinate system, with the origin at each scanning position, to retrieve the corresponding model vector based on the lateral distance between a given voxel and the origin. The matrix-vector multiplications were parallelized and executed on GPU, which greatly improved computational speed.

Numerical simulations

The performance of the suggested MB reconstruction methods, MB-L2 and MB-L1, was first tested on numerical simulations. In the first simulation, the absorption distribution within a (2 x 2 x 3) $mm^3$ FOV was modeled as four truncated paraboloids with diameters 60 μm, 100 μm, 150 μm, and 200 μm (Fig. 1e), and a 25 μm step size was used. The paraboloids were positioned 0.5 mm away from the center of the FOV in the lateral (x and y) directions, and equally separated within a ±1 mm range from the acoustic focus in axial (z) direction. The transducer surface was divided into 1200 sub-elements and the raw OA signals at the center of each sub-element were calculated analytically [70]. The temporal sampling frequency was set to 250 MHz for all simulation experiments, consistent with the experimental setting. The raw signals were normalized to the maximum absolute value and 2% white Gaussian noise (standard deviation of 0.02) was added. MB-L2 and the standard SAFT [10] were used for image reconstruction in this first simulation.

The second simulation was performed by considering two truncated paraboloids with 20 μm diameter at the axial location of the acoustic focus for a FOV of (200 x 200 x 150) $\mu m^3$ and 5 μm step size. The paraboloids were first separated by 60 μm in the x direction to ensure they can be resolved by all three methods in comparison — SAFT, MB-L2, and MB-L1 (Fig. 2a, left). Subsequently, they were separated by 40 µm in the x direction, which was empirically demonstrated to be beyond the acoustic diffraction limit for the filters employed (Fig. 2a, right). Specifically, the raw signals and the model matrix were band-pass filtered between 0.5 MHz and 80 MHz, yielding a diffraction-limited resolution of 61 µm [43]. The raw signal volumes were normalized by the maximum absolute value and no noise was added.

AR-OAM set-up

A recently proposed burst-mode AR-OAM system was used [11]. The system employs a custom-made, 9 µm foil PVDF spherically focused transducer (7 mm focal distance and 0.43 NA). Three laser sources at 532 nm, 578 nm, and 1064 nm were combined and coupled into a multimode fiber, which was guided through a central hole (0.9 mm diameter) of the transducer, thus providing concentric illumination and acoustic detection. Two of the laser sources have fixed wavelengths at 532 nm and 1064 nm (Onda 532 nm and 1064 nm, Bright Solutions, UK), and the third source is a tunable dye laser (Credo, Sirah Lasertechnik, Germany) with wavelength set to 578 nm. The transducer was translated laterally in the x-y plane above the sample (Fig. 1a) to acquire a time-resolved OA signal at each scanning position. Specifically, a fast-moving scanning stage moves constantly forth and back between the boundaries of the previously defined scan window while monitoring its position. Following each pre-defined incremental stage movement, all three lasers were triggered in a cascade with a 6 µs delay in between. Motion artifacts were avoided by averting acceleration and deceleration at each scanning position. The system thus enables rapid acquisition of multi-wavelength volumetric datasets over large FOV. In all experiments, the temporal sampling frequency was set to 250 MSPS and no signal averaging was performed.

Phantom experiments

The suggested MB reconstruction framework was tested on two phantom experiments. A first phantom was used to test if MB-L1 is able to break the acoustic diffraction limit on experimental data. This consisted of a single micro-sphere with 20 µm diameter embedded in 1.3% (w/v) agar. The phantom was placed with the micro-sphere at the axial location of the acoustic focus and a scan covering a (200 x 200 x 150) $\mu m^3$ FOV was performed with 5 µm step size. The 532 nm laser source was used, with up to 2.5 kHz pulse repetition rate (PRR) and 50 µJ per-pulse energy (PPE). The single micro-sphere in the phantom was first individually reconstructed. Subsequently, the signal volume was artificially shifted by 40 µm in the x direction and super-imposed with the original volume, which is equivalent to the signal volume of two micro-spheres separated by the same distance.

A second phantom consisting of four carbon fibers with 7 µm diameter embedded in 1.3% (w/v) agar was imaged. The purpose of this phantom experiment was to assess the overall performance of SAFT and the suggested MB reconstruction methods to reconstruct elongated structures. The scan covered a (1 x 2 x 1) $mm^3$ FOV with 10 µm step size. The PRR was again limited to 2.5 kHz and the PPE of the 532 nm laser was set to 50 µJ.

*In vivo* mouse skin imaging

The *in vivo* imaging performance of the suggested MB reconstruction framework was tested by imaging dorsal mouse skin. For this purpose, a CD-1 mouse (8 weeks old; Charles River Laboratories, Germany) was anesthetized with isoflurane, and placed in prone position on a heating pad. The back skin was shaved and cleaned before imaging. A scan covering a 10 x 30 mm² FOV with a 25 µm step size was performed, giving a volumetric dataset at each wavelength (532 nm, 578 nm, and 1064 nm). The PRR was limited to up to 2.5 kHz and the PPEs for the three lasers were set to 25 µJ, 25 µJ, and 100 µJ, respectively. Before reconstruction, the raw signals were bandpass filtered between 1 MHz and 120 MHz, and further median filtered with kernel size of 3 x 3 x 3. The animal experiment was performed in accordance with the Swiss Federal Act on Animal Protection and were approved by the Canton Veterinary Office Zurich.

Image visualization and vessel quantification

Visualization of the dorsal skin vasculature was done using maximum intensity projections (MIPs) along all Cartesian dimensions. Volume rendering was performed in Paraview 5.8.0 [73]. The images of the mouse skin vasculature were shown with color-coded depth. This was done by multiplying a background map (reconstructed volume), a foreground map (3-channel color encoder), and a transparency value. Foreground colors were set to blue, green, and orange from deeper to shallower regions. The number of vessels was quantified with a previously reported automatic vessel analysis and segmentation (AVSA) algorithm [48]. In short, reconstructed images were binarized, skeletonized, and vessel center-lines fitted by a spline. Vessel edges were found by computing the image gradient along the normal direction of the central line. To quantify the number of vessels achieved in SAFT, MB-L2 and MB-L1, two signal volumes acquired at 532 nm and 578 nm from the multi-wavelength *in vivo* dataset were used. Each of the two volumes were further split into half along the y direction, which increases the sample size and gives a total of four individual datasets. Reconstructed vessels in an upper-right sub-region were individually visualized based on branch point, central line and edge (Fig. 3e). The number of vessels in all four datasets reconstructed with SAFT, MB-L2 and MB-L1 was plotted, and the mean value was noted for each reconstruction method (Fig. 3f).

## Acknowledgments

This work was supported by the Helmut Horten Stiftung (Project Deep Skin, X.L.D.B.) and by the European Research Council under grant agreement ERC-2015-CoG-682379 (D.R.).

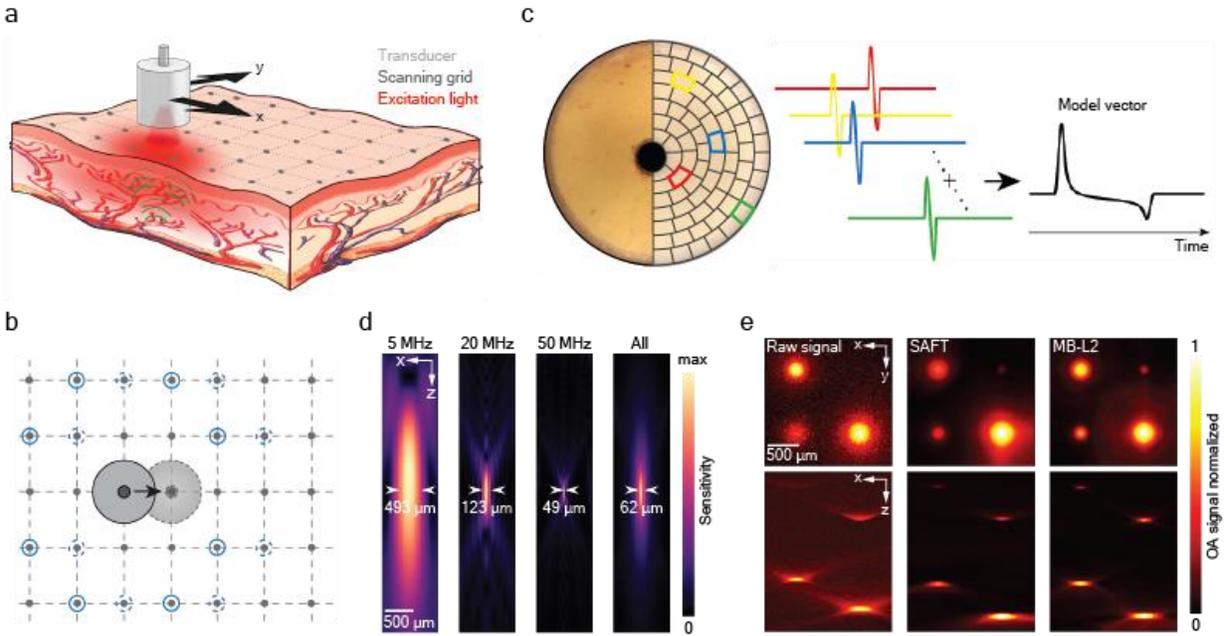

**Fig. 1 Time-domain model for AR-OAM**. (a) Lay-out of the raster-scan scanning protocol. (b) Schematic representation of the scanning symmetries. The OA signals generated at all voxels labeled with blue circles for the indicated transducer position are equivalent. Signals are preserved if the transducer and voxels are shifted by the same distance (dashed lines). (c) Actual photograph of the transducer active area (left) along with a representation of the OA model based on discretizing transducer surface into sub-elements (right). The OA signals corresponding to a given voxel and all surface elements are superimposed to estimate the collected time-resolved signal for that voxel. (d) Cross-sectional view of the amplitude spectral density of the modelled time-resolved signals for three different frequencies along with amplitude of these signals in the time domain. FWHM are indicated. (e) AR-OAM images reconstructed by superimposing the acquired signals (left), with the SAFT method (middle) and with the MB-L2 method (right).

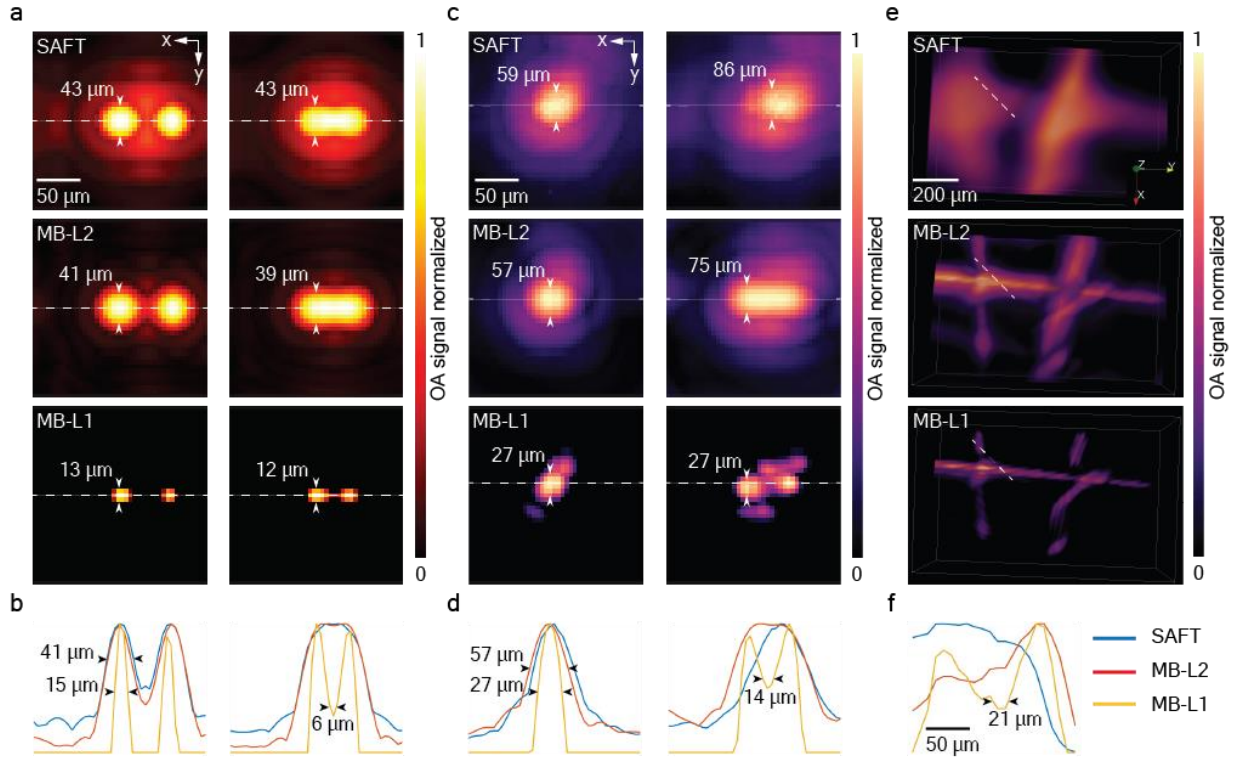

**Fig. 2 Super-resolution imaging beyond the acoustic diffraction limit**. (a) MIPs of the AR-OAM images reconstructed with the SAFT (top), MB-L2 (middle) and MB-L1 (bottom) methods from simulated data corresponding to two 20 μm absorbers separated by 60 μm (left) and 40 μm (right). FWHM are indicated. (b) Cross-sectional profiles for the dashed lines indicated in (a). FWHM are indicated in the left panel, and separation for points >20% of the minimum is indicated in the right panel. (c) MIPs of the AR-OAM images reconstructed with the SAFT (top), MB-L2 (middle) and MB-L1 (bottom) methods from experimental data corresponding to a 20 μm sphere. Reconstructions from the original collected time-resolved signals and from the superposition of signals shifted by 40 μm are shown (left and right respectively). FWHM are indicated. (d) Cross-sectional profiles for the dashed lines indicated in (c). FWHM are indicated in the left panel and separation for points >20% of the minimum is indicated in the right panel. (e) MIPs of the AR-OAM images reconstructed with the SAFT (top), MB-L2 (middle) and MB-L1 (bottom) methods from experimental data corresponding to 7 μm carbon fibers. (f) Cross-sectional profiles for the dashed lines indicated in (e). Separation for points >20% of the minimum is indicated.

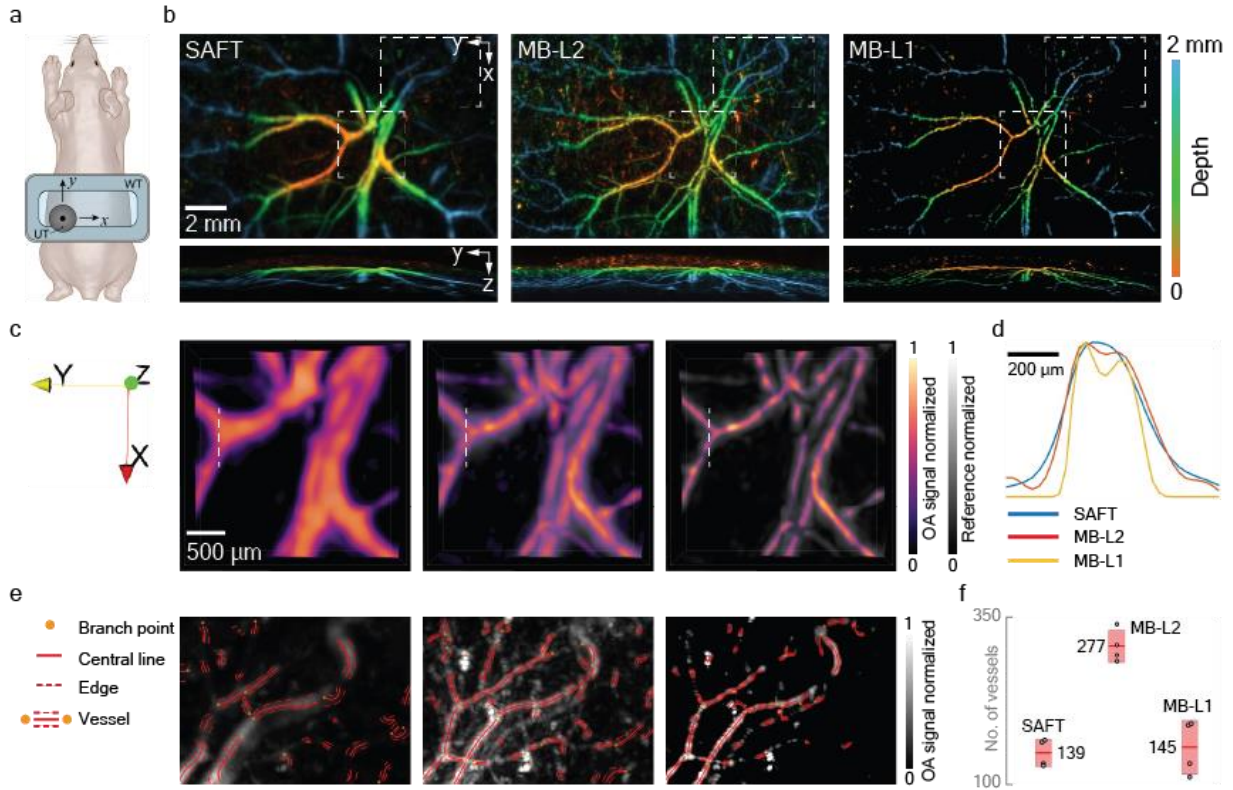

**Fig. 3 Enhanced in vivo imaging capabilities.** (a) Lay-out of the *in vivo* imaging scheme indicating the scanned region. (b) Depth color-coded MIPs of the AR-OAM images reconstructed with the SAFT (left), MB-L2 (middle) and MB-L1 (right) methods. (c) Three-dimensional views of the central region indicated in (b). The reference images are the SAFT image in the middle panel and the MB-L2 image in the right panel. (d). Cross-sectional profiles indicated in (c). (e) Reconstructed vessels in the binarized image corresponding to the indicated region in the upper right corner in (b). (f) Number of detected vessels for four different *in vivo* datasets reconstructed with SAFT, MB-L2 and MB-L1. The average number of vessels are indicated for each reconstruction method.